\documentclass[letterpaper, 10 pt, conference]{ieeeconf}  

\IEEEoverridecommandlockouts                              

\usepackage{amsmath} 
\usepackage{amssymb}  
\usepackage{cite}

\usepackage{graphicx}

\newtheorem{theorem}{Theorem}

\newtheorem{definition}{Definition}
\newtheorem{remark}{Remark}

\title{\LARGE \bf
Switching control for incremental stabilization of nonlinear systems via contraction theory
}

\author{Mario di Bernardo$^{1,2}$ and Davide Fiore$^{1}$
\thanks{$^{1}$M. di Bernardo and D. Fiore are with the Department of Electrical Engineering and Information Technology, University of Naples Federico II, Via Claudio 21, 80125 Naples, Italy
        {\tt\small davide.fiore@unina.it}}%
\thanks{$^{2}$M. di Bernardo is also with the Department of Engineering Mathematics, University of Bristol, University Walk, BS8 1TR Bristol, U.K.
        {\tt\small mario.dibernardo@unina.it}}%
}

\begin{document}

\maketitle
\thispagestyle{empty}
\pagestyle{empty}

\begin{abstract}
In this paper we present a switching control strategy to incrementally stabilize a class of nonlinear dynamical systems. Exploiting recent results on contraction analysis of switched Filippov systems derived using regularization, sufficient conditions are presented to prove incremental stability of the closed-loop system. Furthermore, based on these sufficient conditions, a design procedure is proposed to design a switched control action that is active only where the open-loop system is not sufficiently incrementally stable in order to reduce the required control effort. The design procedure to either locally or globally incrementally stabilize a dynamical system is then illustrated by means of a representative example.
\end{abstract}

\section{Introduction}
Incremental stability has been established as a powerful tool to prove convergence in nonlinear dynamical systems \cite{angeli2002lyapunov}. An effective approach to obtain sufficient conditions for incremental stability comes from contraction theory \cite{lohmiller1998contraction,russo2010global,jouffroy2005some, forni2014differential,aminzare2014contraction}. More specifically, incremental exponential stability over a given forward invariant set is guaranteed if some matrix measure $\mu$ of the system Jacobian matrix is uniformly negative in that set for all time. Moreover, contraction theory has been used as a synthesis tool to design incrementally stabilizing controllers and observers \cite{lohmiller1998contraction,manchester2014control,manchester2014output,van2008tracking}.

Piecewise smooth (PWS) systems are important in applications, ranging from problems in mechanics (friction, impact) and biology (genetic regulatory networks) to variable structure systems in control engineering \cite{filippov1988differential,cortes2008discontinuous,bernardo2008piecewise,utkin2013sliding}. Several results have been presented in literature to extend contraction analysis to these classes of nondifferentiable vector field \cite{lohmiller2000nonlinear,pavlov2005convergentp1,pavlov2005convergentp2,di2014contraction,lu2015contraction,di2013incremental,di2014incremental,fiore2015contraction}.

In this paper we discuss the problem of designing a switched feedback control to incrementally stabilize a nonlinear dynamical systems over some set of interest. Our approach is based on some of our previous analytical results on contraction and incremental stability of bimodal Filippov systems which were recently presented in \cite{fiore2015contraction}. In particular the switching control action resulting from our design procedure is active only where the open-loop system is not sufficiently incrementally stable. Such behavior can be usefully exploited to reduce the required control effort.

The rest of the paper is organized as follows. Section \ref{sec:background} summarizes the necessary mathematical preliminaries on contraction analysis and incremental stability of continuously differentiable systems and recalls a basic result on bimodal Filippov systems presented in \cite{fiore2015contraction}. Section \ref{sec:control_design} contains our main result on switched controlled systems and a design procedure to derive an incrementally stabilizing switching control input. The design procedure is illustrated with an example in Section \ref{sec:examples}. Conclusions are drawn in Section \ref{sec:conclusions}.


\section{Contraction analysis of PWS systems}
\label{sec:background}
\subsection{Incremental Stability and Contraction Theory}
Let $U\subseteq\mathbb{R}^n$ be an open set. Consider the system of ordinary differential equations
\begin{equation}
\label{eq:dynamical_sys}
\dot{x}=f(t,x)
\end{equation}
where $f$ is a continuously differentiable vector field defined for $t\in[0,\infty)$ and $x\in U$, that is $f\in C^1(\mathbb{R}^+\times U,\mathbb{R}^n)$.

We denote by $\psi(t,t_0,x_0)$ the value of the solution $x(t)$ at time $t$ of the differential equation \eqref{eq:dynamical_sys} with initial value $x(t_0)=x_0$.
We say that a set $\mathcal{C}\subseteq \mathbb{R}^n$ is \emph{forward invariant} for system \eqref{eq:dynamical_sys}, if $x_0 \in \mathcal{C}$ implies $\psi(t,t_0,x_0) \in \mathcal{C}$ for all $t\ge t_0$.

\begin{definition}
\label{def:incr_stability}
Let $\mathcal{C}\subseteq\mathbb{R}^n$ be a forward invariant set and $|\cdot|$ some norm on $\mathbb{R}^n$. The system \eqref{eq:dynamical_sys} is said to be \emph{incrementally exponentially stable} ($IES$) in $\mathcal{C}$ if there exist constants $K\geq 1$ and $\lambda>0$ such that 
\begin{equation}
\label{eq:ies}
\lvert x(t)-y(t)\rvert \leq  K\, e^{-\lambda(t-t_0)}\,\lvert x_0-y_0\rvert 
\end{equation}
$\forall t \geq t_0$, $\forall x_0,y_0 \in \mathcal{C}$, where $x(t)=\psi(t,t_0,x_0)$ and $y(t)=\psi(t,t_0,y_0)$ are its two solutions.
\end{definition}

Results in contraction theory can be applied to a quite general class of subsets $\mathcal{C}\subseteq\mathbb{R}^n$, known as K-reachable subsets \cite{russo2010global}. See Appendix for a definition.

\begin{definition}
\label{def:contraction}
The continuously differentiable vector field \eqref{eq:dynamical_sys} is said to be \emph{contracting} on a K-reachable set $\mathcal{C}\subseteq U$ if there exists some norm in $\mathcal{C}$, with associated matrix measure $\mu$ (see Appendix), such that, for some constant $c>0$ (the \emph{contraction rate})
\begin{equation}
\label{eq:contraction_cond}
\mu\left(\frac{\partial f}{\partial x}(t,x)\right)\leq -c, \quad \quad \forall x\in\mathcal{C},\quad \forall t\geq t_0.
\end{equation}
\end{definition}
\vspace{0.2cm}
The basic result of nonlinear contraction analysis states that, if a system is contracting, then all of its trajectories are incrementally exponentially stable, as follows.
\begin{theorem}
\label{thm:contraction}
Suppose that $\mathcal{C}$ is a K-reachable forward-invariant subset of $U$ and that the vector field \eqref{eq:dynamical_sys} is infinitesimally contracting with contraction rate $c$ therein. Then, for every two solutions $x(t)=\psi(t,t_0,x_0)$ and $y(t)=\psi(t,t_0,y_0)$ with $x_0,y_0\in\mathcal{C}$ we have that \eqref{eq:ies} holds with $\lambda=c$.
\end{theorem}
As a result, if a system is contracting in a forward-invariant subset then it converges towards an equilibrium point therein \cite{russo2010global,lohmiller1998contraction}.

In this paper we analyse contraction properties of dynamical systems based on norms and matrix measures \cite{russo2010global}. Other more general definitions exist in the literature, for example results based on Riemannian metrics \cite{lohmiller1998contraction} and Finsler-Lyapunov functions \cite{forni2014differential}. The relations between these three definitions and the definition of convergence \cite{pavlov2004convergent} have been investigated in \cite{forni2014differential}.

\subsection{Filippov systems}
The control input $u(x)$ we are going to design in this paper is a discontinuous function, this implies that even if the open-loop vector field is continuously differentiable the resulting closed-loop vector field is obviously not. In particular it belongs to a class of systems that has been investigated by Filippov \cite{filippov1988differential} and Utkin \cite{utkin2013sliding}. Switched (or bimodal) Filippov systems are dynamical systems $\dot{x}=f(x)$ where $f(x)$ is a piecewise continuous vector field having a codimension-one submanifold $\Sigma$ as its discontinuity set.
 
\begin{figure}[t]
\centering 
{\includegraphics[width=2.5in]{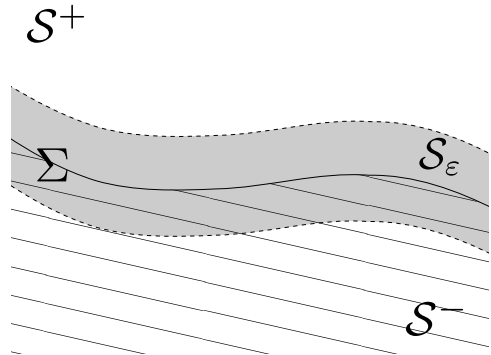}}
\caption{Regions of state space: the switching manifold, $\Sigma:=\{x\in U : H(x)=0\}$ from \eqref{eq:switching_manifold}, 
$\mathcal{S}^+:=\{x\in U : H(x)>0\}$, $\mathcal{S}^-:=\{x\in U : H(x)<0\}$ (hatched zone) and $\mathcal{S}_{\varepsilon}:=\{x\in U : -\varepsilon <H(x)<\varepsilon \}$ (grey zone) from \eqref{eq:regularizationregion}.} 
\label{fig:regions}
\end{figure}

The submanifold $\Sigma$ is called the \emph{switching manifold} and is defined as the zero set of a smooth function $H:\,U\rightarrow\mathbb{R}$, that is
\begin{equation}
\label{eq:switching_manifold}
\Sigma:=\{x\in U : H(x)=0\} 
\end{equation}
where $0\in\mathbb{R}$ is a regular value of $H$, i.e. $\nabla H(x)\neq 0,\, \forall x\in\Sigma$. It divides $U$ in two disjoint regions, $\mathcal{S}^+:=\{x\in U : H(x)>0\}$ and ${\mathcal{S}^-:=\{x\in U : H(x)<0\}}$ (see Figure \ref{fig:regions}).

Hence, a bimodal Filippov system can be defined as
\begin{equation}
\label{eq:filippov_bimodal}
\dot{x}=
\begin{cases}
F^+(x) \quad \mbox{if } x\in\mathcal{S}^+ \\
F^-(x) \quad \mbox{if } x\in\mathcal{S}^-
\end{cases}
\end{equation}
where $F^+,F^-\in{C}^1(U,\mathbb{R}^n)$.  When the normal components of the vector fields either side of $\Sigma$ point in the \textit{same} direction, the gradient of a trajectory is discontinuous, leading to Carath\'eodory solutions \cite{filippov1988differential}. In this case, the dynamics is described as \textit{crossing} or \textit{sewing}. But when the vector fields on either side of $\Sigma$ both point toward it, the solutions are constrained to evolve along $\Sigma$ and some additional dynamics needs to be given when such \emph{sliding} behavior occurs. To define this sliding vector field it is widely adopted the Filippov convention \cite{filippov1988differential}. 
\begin{remark}
In the following we assume that solutions of system \eqref{eq:filippov_bimodal} are defined in the sense of Filippov \cite{filippov1988differential} and they have the property of \emph{right-uniqueness} \cite[pag. 106]{filippov1988differential} holds in $U$, i.e. for each point $x_0\in U$ there exists $t_1>t_0$ such that any two solutions satisfying $x(t_0)=x_0$ coincide on the interval $[t_0,\, t_1]$. Therefore, the escaping region is excluded from our analysis. 
\end{remark}

Definition \ref{def:contraction} was previously presented as a sufficient condition for a dynamical system to be incrementally exponentially stable, but  condition \eqref{eq:contraction_cond} can not be directly applied to system \eqref{eq:filippov_bimodal} because its vector field is not continuously differentiable. Therefore an extension of contraction analysis to PWS systems is not straightforward. In a recent work \cite{fiore2015contraction} sufficient conditions were derived for convergence of any two trajectories of a Filippov system between each other. Instead of directly analyzing the Filippov system, a regularized version $f_\varepsilon(x)$ was considered given as
\begin{equation*}
\label{eq:regularized_sys}
f_\varepsilon(x)=\frac{1+\varphi_\varepsilon \left( H(x) \right)}{2}\, F^+(x) +
\frac{1-\varphi_\varepsilon \left( H(x) \right)}{2}\, F^-(x)
\end{equation*}
where $\varphi_\varepsilon\in C^1(\mathbb{R},\mathbb{R})$ is the so-called transition function. See the original paper \cite{sotomayor1996regularization} from Sotomayor and Teixeira for further details on the regularization method adopted in \cite{fiore2015contraction}. 
In this new system the switching manifold $\Sigma$ has been replaced by a boundary layer $\mathcal{S}_\varepsilon$ (Figure \ref{fig:regions}) of width $2\varepsilon$
\begin{equation}
\label{eq:regularizationregion}
\mathcal{S}_{\varepsilon}:=\{x\in U : -\varepsilon <H(x)<\varepsilon \}
\end{equation}
and more important $f_\varepsilon$ is continuously differentiable in $U$, therefore condition \eqref{eq:contraction_cond} can be applied to it. Finally, results that are valid for Filippov systems \eqref{eq:filippov_bimodal} were recovered taking the limit for $\varepsilon\rightarrow 0$.

The sufficient conditions for a bimodal Filippov system to be incrementally exponentially stable in a certain set are stated in the following theorem from \cite{fiore2015contraction}. 

\begin{theorem}
\label{thm:contracting_pws}
The bimodal Filippov system \eqref{eq:filippov_bimodal} is incrementally exponentially stable in a K-reachable set $\mathcal{C}\subseteq U$ with convergence rate $c:=\min\,\{c_1,c_2\}$ if there exists some norm in $\mathcal{C}$, with associated matrix measure $\mu$, such that for some positive constants $c_1,c_2$
\begin{eqnarray*}
\label{eq:thm:condition1}
\mu\left( \frac{\partial F^+}{\partial x}(x)\right) \leq -c_1, &\forall x \in \bar{\mathcal{S}}^+\\
\label{eq:thm:condition2}
\mu\left( \frac{\partial F^-}{\partial x}(x)\right) \leq -c_2,  &\forall x \in \bar{\mathcal{S}}^-\\ 
\label{eq:thm:condition3}
\mu\left( \Big[ F^+(x)-F^-(x)\Big] \cdot \nabla H(x) \right) = 0,  &\forall x \in \Sigma.
\end{eqnarray*}
\end{theorem}
\vspace{0.3cm}
In the above relations $\bar{\mathcal{S}}^+$ and $\bar{\mathcal{S}}^-$ represent the closures of the sets $\mathcal{S}^+$ and $\mathcal{S}^-$, respectively.
The interested reader can refer to \cite{fiore2015contraction} for a complete proof and further details.

\section{Switching control design}
\label{sec:control_design}
\subsection{Problem formulation}
In this paper we consider the class of dynamical systems defined by
\begin{equation}
\label{eq:controlled_sys}
\dot{x}=f(x)+g(x)\,u(x)
\end{equation}
where $x\in\mathbb{R}^n$, $u(x)\in\mathbb{R}^m$ are state and feedback control input, and  $f:\mathbb{R}^n\rightarrow\mathbb{R}^n$, $g:\mathbb{R}^n\rightarrow\mathbb{R}^{n\times m}$ are continuously differentiable vector fields.

We want to find a discontinuous feedback control input $u$ for system \eqref{eq:controlled_sys} such that the resulting closed-loop system is incrementally stabilized, either locally or globally. The control input $u(x)$ we are looking for has the following form
\begin{equation}
\label{eq:control_input}
u(x)=
\begin{cases}
u^+(x) & \mbox{if } H(x)>0\\
u^-(x) & \mbox{if } H(x)<0
\end{cases}
\end{equation}
where $u^\pm(x)$ are continuously differentiable vector fields, and $H(x)$ is a scalar function as in \eqref{eq:switching_manifold}.

In particular, to minimize the control effort we want to exploit possible contracting properties of the open-loop vector field $f(x)$ to design a control input that is not active in the regions where $f(x)$ is already sufficiently incrementally stable.

\subsection{Main theorem}
The main result of this paper follows directly from Theorem \ref{thm:contracting_pws}.
\begin{theorem}
\label{thm:main_control}
The dynamical systems \eqref{eq:controlled_sys} with the switching control input \eqref{eq:control_input} is incrementally exponentially stable in a K-reachable set $\mathcal{C}\subseteq U$ with convergence rate $c:=\min\,\{c_1,c_2\}$ if there exist some norm in $\mathcal{C}$, with associated matrix measure $\mu$ such that for some positive constants $c_1, c_2$
\begin{eqnarray}
\label{eq:thm:control_condition1}
\mu\left(\frac{\partial f}{\partial x}(x)+\frac{\partial}{\partial x}\big[ g(x)\,u^+(x) \big] \right) \leq -c_1, &\forall x \in \bar{\mathcal{S}}^+\\
\label{eq:thm:control_condition2}
\mu\left( \frac{\partial f}{\partial x}(x)+\frac{\partial}{\partial x}\big[ g(x)\,u^-(x) \big]\right) \leq -c_2, &\forall x \in \bar{\mathcal{S}}^-\\ 
\label{eq:thm:control_condition3}
\mu\Big( g(x)\,\big[ u^+(x)-u^-(x)\big] \cdot \nabla H(x) \Big) = 0, &\forall x \in \Sigma
\end{eqnarray}
\end{theorem}
\vspace{0.3cm}
\begin{proof}
The closed-loop system with switching control \eqref{eq:control_input} is a Filippov system as \eqref{eq:filippov_bimodal} of the form
\begin{equation}
\label{eq:closed_loop_sys}
\dot{x}=
\begin{cases}
F^+(x):=f(x)+g(x)\,u^+(x) & \mbox{if } H(x)>0\\
F^-(x):=f(x)+g(x)\,u^-(x) & \mbox{if } H(x)<0
\end{cases}
\end{equation}
therefore Theorem \ref{thm:contracting_pws} can be directly applied giving the previous three conditions. And thus if these conditions hold then the switching control \eqref{eq:control_input} incrementally stabilizes system \eqref{eq:controlled_sys} with convergence rate $c$.
\end{proof}

Note that
\begin{equation*}
\frac{\partial}{\partial x}\left[ g(x)\,u^\pm(x) \right]=
\sum_{i=1}^m\left( \frac{\partial g_i}{\partial x}(x)\,u^\pm_i(x)+g_i(x)\,\frac{\partial u^\pm_i}{\partial x}(x)  \right)
\end{equation*}
where we denoted with $g_i$ and $u^\pm_i$ the $i$-th column of $g(x)$ and the $i$-th component of $u^\pm(x)$, respectively.

\subsection{Design procedure}

In the following we present a possible approach to design a switching controller \eqref{eq:control_input} that incrementally stabilize system \eqref{eq:controlled_sys} in a desired set using conditions of Theorem \ref{thm:main_control}. Indeed if the designed $u(x)$ is such that conditions \eqref{eq:thm:control_condition1}-\eqref{eq:thm:control_condition3} are satisfied for a desired $c$ then the discontinuous closed-loop system \eqref{eq:closed_loop_sys} is incrementally exponentially stable as required.

Specifically, suppose that it is required for the closed-loop system \eqref{eq:closed_loop_sys} to be incrementally stable with convergence rate $\bar{c}$ in a certain set $\mathcal{C}_d$ (where the open-loop system \eqref{eq:dynamical_sys} is not sufficiently contracting).

Suppose that in $\mathcal{C}_d$ there can be identified two disjoint subregions, one where condition \eqref{eq:contraction_cond} with $c=\bar{c}$ is not satisfied and the other one where it is satisfied (without the equality sign). Specifically, the two subregions are 
\begin{eqnarray*}
\mathcal{S}^+:= \left\{ x\in\mathcal{C}_d:\, \mu\left( \frac{\partial f}{\partial x}(x) \right)>-\bar{c} \right\}\\
\mathcal{S}^-:= \left\{ x\in\mathcal{C}_d:\, \mu\left( \frac{\partial f}{\partial x}(x) \right)<-\bar{c} \right\}
\end{eqnarray*}

The key design idea is to choose the scalar function $H$ in \eqref{eq:control_input} as
\begin{equation}
\label{eq:H_control}
H(x)=\mu\left(\frac{\partial f}{\partial x}(x)\right)+\bar{c},
\end{equation}
in this way the switching manifold $\Sigma$ is defined as 
\begin{equation}
\label{eq:Sigma_control}
\Sigma := \left\{ x\in\mathcal{C}_d:\, \mu\left(\frac{\partial f}{\partial x}(x)\right)=-\bar{c} \right\}.
\end{equation}

The final step is to find $u^+$ and $u^-$ such that conditions \eqref{eq:thm:control_condition1}-\eqref{eq:thm:control_condition3} are satisfied. Obviously with the selection of $H(x)$ made in \eqref{eq:H_control} the open-loop vector field $f$ already satisfies the design requirements in $\mathcal{S}^-$, therefore in this case the simplest choice is 
\begin{equation}
u^-(x)=0,
\end{equation}
and the control problem is reduced to find a $u^+$ that satisfies \eqref{eq:thm:control_condition1} and \eqref{eq:thm:control_condition3}. In other terms, by selecting \eqref{eq:Sigma_control} as switching manifold the resulting switching control input can be active only in the region where the controlled system is not sufficiently contracting. 

This property can be exploited to reduce the average control energy compared to the one required by a continuous control input defined in the whole set $\mathcal{C}_d$ (eventually globally), as we will show in the next section through a simple example.

\section{Representative examples}
\label{sec:examples}

Here we present examples to illustrate the design procedure described in the previous section. The unweighted 1-norm will be used to highlight that non-Euclidean norms can be used in some cases as an alternative to Euclidean norms and that not only the analysis but the control synthesis too can be easier if they are used. See Appendix for the definition of the matrix measure induced by unweighted 1-norm $\mu_1$.

The nonlinear system \eqref{eq:controlled_sys} that we want to incrementally stabilize in a certain set is 
\begin{equation}
\label{eq:example_open_loop}
\dot{x}=
\begin{bmatrix}
-4x_1\\
x_2^2-6x_2
\end{bmatrix}
+
\begin{bmatrix}
1\\
2
\end{bmatrix}
u(x)
\end{equation}

The desired convergence rate $\bar{c}$ in this examples is set to 2, i.e. $\bar{c}=2$.

It can be easily seen that
\begin{equation*}
\begin{split}
\mu_1 \left(\frac{\partial f}{\partial x}(x)\right) &=
\mu_1\left( 
\begin{bmatrix}
-4 & 0\\
0 & 2x_2-6
\end{bmatrix} 
\right)=\\
&= \max\,\{ -4;\; 2x_2-6\}=\\
&=\begin{cases}
-4 & \mbox{if } x_2\leq 1\\
2x_2-6 & \mbox{if } x_2>1
\end{cases}
\end{split}
\end{equation*}
Therefore the set $\mathcal{C}$ where system \eqref{eq:example_open_loop} is contracting with contraction rate $\bar{c}$, that is where it satisfies condition \eqref{eq:contraction_cond}, is  
\begin{equation*}
\mathcal{C}=\{x\in\mathbb{R}^2:\, x_2<2\}.
\end{equation*}

In the following two design examples will be presented and discussed. In the first one we want to extend the region $\mathcal{C}$ where the system is incrementally stable to the set $\mathcal{C}_d\supset\mathcal{C}$, and in the second one we want to make the system globally incrementally stable, that is $\mathcal{C}_d\equiv\mathbb{R}^2$.

In both cases, following the design procedure of Section \ref{sec:control_design}, the scalar function $H$ of the switching controller is set as
\begin{equation*}
H(x)=\mu_1\left(\frac{\partial f}{\partial x}(x)\right)+2
\end{equation*}
and the switching manifold $\Sigma$ as its zero set, that is as
\begin{equation*}
\Sigma=\left\{ x\in\mathcal{C}_d:\; x_2=2 \right\}
\end{equation*}

Furthermore, as expected the control requirements are already satisfied in $\mathcal{S}^-$, and thus $u^-(x)=0$. The problem is now reduced to find a function $u^+(x)$ such that conditions \eqref{eq:thm:control_condition1} and \eqref{eq:thm:control_condition3} hold. Specifically, condition \eqref{eq:thm:control_condition1} is satisfied if the following quantity is made less than $-\bar{c}$
\begin{equation}
\label{eq:example_mu1}
\begin{split}
\mu_1& \left(\frac{\partial}{\partial x}\big[ f(x)+g(x)\,u^+(x) \big] \right)=\\
=&\mu_1\left( 
\begin{bmatrix}
-4+u_{x1} & u_{x2}\\
2u_{x1} & 2x_2-6+2u_{x2}
\end{bmatrix} 
\right)=\\
=& \max\,\{ -4+u_{x1}+\lvert 2u_{x1} \rvert;\\
& \qquad \quad 2x_2-6+2u_{x2} +\lvert u_{x2}\rvert\}
\end{split}
\end{equation}
with $\frac{\partial u^+}{\partial x}=[u_{x1}\;\; u_{x2}]$.\\
In this simple example the first term in \eqref{eq:example_mu1} does not depend on $x$ so it can be made less than $-\bar{c}$ by simply setting $u_{x1}=0$. Therefore, in conclusion we need to find $u_{x2}$ such that
\begin{equation}
\label{eq:example_design_cond}
2x_2-6+2u_{x2}+|u_{x2}|\leq-2, \qquad \forall x\in\bar{\mathcal{S}}^+
\end{equation}
and then check if the resulting $u(x)$ satisfies \eqref{eq:thm:control_condition3} where $\nabla H=[0\;\; 1]$.

\subsection{Example 1}
\label{sec:example_1}
As previously said, we want to extend the region where system \eqref{eq:example_open_loop} is contracting to a new set $\mathcal{C}_d$, in particular we choose $\mathcal{C}_d=\{x\in\mathbb{R}^2:\, x_2<7\}$. Therefore $\mathcal{S}^+=\{x\in\mathcal{C}_d:\; 2<x_2<7\}$, and it can be easily proved that \eqref{eq:example_design_cond} is satisfied for $u_{x2}\leq-10$, and thus, by integration, we have 
\begin{equation*}
u^+(x)=-10x_2
\end{equation*}

Condition \eqref{eq:thm:control_condition3} is also satisfied, since we have that for all $x\in\Sigma$
\begin{equation*}
\begin{split}
\mu_1&\left(
\begin{bmatrix}
1\\
2
\end{bmatrix}
\cdot (-10x_2) \cdot
\begin{bmatrix}
0 & 1
\end{bmatrix} \right)=\\
=& 10\,\mu_1\left( 
\begin{bmatrix}
0 & -x_2\\
0 & -2x_2
\end{bmatrix} 
\right)=\\
=& 10\, \max\,\{ 0;\; -2x_2+|x_2|\}=\\
=& 10\, \max\,\{ 0;\; -2\}=0.
\end{split}
\end{equation*}

\begin{figure}[t]
\centering 
{\includegraphics[width=3.2in]{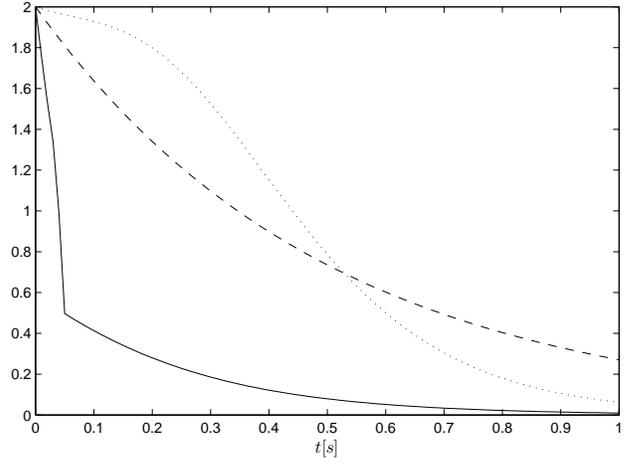}}
\caption{System \eqref{eq:example_open_loop} in open-loop (dotted line) and with control \eqref{eq:example1_control} (solid line). Initial conditions in $x_0=[1\;\;4]^T$ and $y_0=[2\;\;5]^T$. The dashed line is the estimated exponential decay from \eqref{eq:ies} with $\lambda=\bar{c}=2$ and $K=1$. } 
\label{fig:example1_sim}
\end{figure}

In conclusion a switching control input that incrementally stabilize \eqref{eq:example_open_loop} in $\mathcal{C}_d$ is
\begin{equation}
\label{eq:example1_control}
u(x)=
\begin{cases}
-10x_2 & \mbox{if } x_2>2\\
0 & \mbox{if } x_2<2
\end{cases}
\end{equation}
In Figure \ref{fig:example1_sim}, we report numerical simulations of the evolution of the difference between two trajectories. The dashed line is the estimated decay from \eqref{eq:ies} with $c=2$ and $K=1$. It can be seen that as expected
\begin{equation*}
\lvert x(t)-y(t)\rvert_1\leq e^{-2t}\,\lvert x_0-y_0\rvert_1, \qquad \forall t>0.
\end{equation*}

\subsection{Example 2}
\label{sec:example_2}
If we want system \eqref{eq:example_open_loop} to be globally incrementally stable (that is $\mathcal{C}_d\equiv\mathbb{R}^2$) condition \eqref{eq:example_design_cond} has to be verified with $\mathcal{S}^+=\{x\in\mathbb{R}^2:\; x_2>2\}$. It can be proved that such condition is satisfied choosing for example $u_{x2}=-2x_2$, and therefore by integration the control input defined in $\mathcal{S}^+$ is
\begin{equation*}
u^+(x)=-x_2^2.
\end{equation*}
%

Again, condition \eqref{eq:thm:control_condition3} is satisfied since
\begin{equation*}
\begin{split}
\mu_1&\left(
\begin{bmatrix}
1\\
2
\end{bmatrix}
\cdot (-x_2^2) \cdot
\begin{bmatrix}
0 & 1
\end{bmatrix} \right)=\\
=& \max\,\{ 0;\; -2x_2^2+|-x_2^2|\}=\\
=& \max\,\{ 0;\; -4\}=0
\end{split}
\end{equation*}
for all $x\in\Sigma$.

\begin{figure}[t]
\centering 
{\includegraphics[width=3.2in]{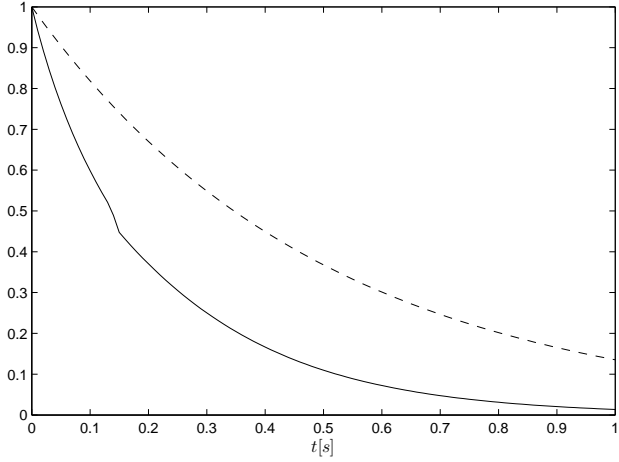}}
\caption{Closed-loop system with control \eqref{eq:example2_control}. Initial conditions in $x_0=[1\;\;8]^T$ and $y_0=[1\;\;9]^T$. The dashed line is the estimated exponential decay from \eqref{eq:ies} with $\lambda=\bar{c}=2$ and $K=1$.} 
\label{fig:example2_sim}
\end{figure}

To conclude, system \eqref{eq:example_open_loop} is globally incrementally stabilized by the switching controller
\begin{equation}
\label{eq:example2_control}
u(x)=
\begin{cases}
-x_2^2 & \mbox{if } x_2>2\\
0 & \mbox{if } x_2<2
\end{cases}
\end{equation}
In Figure \ref{fig:example2_sim}, we show numerical simulations of the evolution of the difference between two trajectories that confirm the theoretical results. Open-loop simulations are not reported in this case since the system is unstable for chosen initial conditions.

All simulations presented in this section were computed using the numerical solver in \cite{piiroinen2008event}.

\subsection{Discussion}
As highlighted in the previous section, the control inputs designed here are active only in the region $\mathcal{S}^+$ of the state space where the open-loop system is not sufficiently incrementally stable, otherwise they are turned off. On the other hand, to satisfy the same stability requirements a continuous control feedback $\widehat{u}(x)$ has to be design such that
\begin{equation*}
\mu \left(\frac{\partial}{\partial x}\big[ f(x)+g(x)\,\widehat{u}(x) \big] \right)\leq -\bar{c} \qquad \forall x\in\mathcal{C}_d,
\end{equation*}
and thus it has to take non-zero values on the whole $\mathcal{C}_d$.
Therefore, the switching control law presented in this paper has the additional property that it can be turned off in $\mathcal{S}^-$ to reduce the required control energy.

For example, a continuous feedback control that satisfies control requirements as in Example 1 is
\begin{equation}
\label{eq:example1_smooth_control}
\widehat{u}(x)=-10x_2 \qquad \forall x\in\mathcal{C}_d
\end{equation}
that is $u^+(x)$ in \eqref{eq:example1_control} extended to $\mathcal{S}^-$. Hence in this case it is clear that control input \eqref{eq:example1_control} uses less energy than \eqref{eq:example1_smooth_control}.

Instead, for what concerns Example 2, a continuous function $\widehat{u}(x)$ such that \eqref{eq:example_design_cond} holds on all $\mathbb{R}^2$ has to be at least cubic (while \eqref{eq:example2_control} is quadratic). Since their derivatives have to satisfy the same linear constraint \eqref{eq:example_design_cond} in $\mathcal{S}^+$, it easily follows that the $L_2$-norm of the continuous control input will be always greater than the one of the discontinuous input. 
%


\section{Conclusions}
\label{sec:conclusions}

In this paper we formulated the problem of designing a switched control action to stabilize a nonlinear system using tools from contraction theory. Based on sufficient conditions for incremental exponential stability of switched bimodal Filippov system derived in \cite{fiore2015contraction}, we presented a control design strategy to incrementally stabilize a class of nonlinear systems. The effectiveness of the design methodology to derive both global and local results was illustrated through a simple but representative example. Moreover, we showed that different metrics rather than the Euclidean norm can be effectively used in the design of the controller.

Future work will be aimed at extending the class of systems stabilizable through such switched controllers and to construct state observers for these systems using methodologies similar to those presented here. Furthermore, it is of interest to reformulate the design procedure as a convex optimization problem to compute numerically both metrics and control gains.


\addtolength{\textheight}{-9.5cm}   



\section*{Appendix}
\subsection*{K-reachable sets}
Let $K>0$ be any positive real number. A subset $\mathcal{C}\subseteq\mathbb{R}^n$ is \emph{K-reachable} if for any two points $x_0$ and $y_0$ in $\mathcal{C}$ there is some continuously differentiable curve $\gamma:[0,1]\rightarrow\mathcal{C}$ such that $\gamma(0)=x_0$, $\gamma(1)=y_0$ and $\lvert \gamma'(r)\rvert \leq K\lvert y_0-x_0\rvert,\; \forall\, r$.

For convex sets $\mathcal{C}$, we may pick $\gamma(r)=x_0+r(y_0-x_0)$, so $\gamma'(r)=y_0-x_0$ and we can take $K=1$. Thus, convex sets are 1-reachable, and it is easy to show that the converse holds.

\subsection*{Matrix measure}
The \emph{matrix measure} \cite{vidyasagar2002nonlinear} associated to a matrix $ A \in\mathbb{R}^{n \times n}$ is the function $\mu(\cdot):\mathbb{R}^{n \times n}\rightarrow \mathbb{R}$ defined as
\begin{equation*}
\label{eq:matrix_measure}
\mu( A )=\lim_{h \rightarrow 0^+}\frac{\lVert I+h A  \rVert-1}{h}
\end{equation*}
The measure of a matrix $ A $ can be thought of as the one-sided directional derivative of the induced matrix norm function $\lVert \cdot \rVert$, evaluated at the point $I$, in the direction of $ A $. See \cite{vidyasagar1978matrix} for a more general definition of matrix measure induced by a positive convex function and \cite{vidyasagar2002nonlinear,desoer1972measure} for a list of properties of this measure.

In this paper we often use the measure induced by unweighted 1-norm:
\begin{equation*}
\label{eq:mu_1}
\mu_{1}( A )=\max_{j}\left[a_{jj}+\sum_{i\ne j}|a_{ij}|\right].
\end{equation*}
Other matrix measures often used in literature are the one induced by Euclidean norm
$$
\mu_{2}( A )=\lambda_{max}\left(\frac{ A + A^T }{2}\right),
$$ 
and the one induced by $\infty$-norm
$$
\mu_{\infty}( A )=\max_{i}\left[a_{ii}+\sum_{j\ne i}|a_{ij}|\right].
$$
\vspace{0.1cm}
%
%
%


\bibliographystyle{IEEEtranBST/IEEEtran} 
\bibliography{IEEEtranBST/IEEEabrv,refs}

\end{document}